**Decoding Magnetic Texture**

*Michael P. Path[1]*, Jeffrey McCord[1,2], Michael Vogel[1,2]**

[1]Nanoscale Magnetic Materials - Magnetic Domains, Department of Materials Science, Kiel University, 24143 Kiel, Germany

[2]Kiel Nano Surface and Interface Science (KiNSIS), Kiel University, 24143 Kiel, Germany

E-mail: mipa@tf.uni-kiel.de, mv@tf.uni-kiel.de

Funding: German Research Foundation (DFG) - MC 9/20-2; VO 3017/1-1

Keywords: magnetic materials, magnetic textures, magnetic domains, magneto-optics, machine learning, deep learning

## Abstract:

In magnetically ordered materials, magnetic field and temperature variations modify the magnetic texture through their coupling to the local energy landscape, imprinting distinct fingerprints in the resulting magnetic domain patterns. Retrieving these conditions from the pattern remains challenging, as stochastic nucleation and hysteresis produce a nonlinear, multivariate, and ambiguous relationship between magnetic domain morphology and external stimuli. To decode these fingerprints, we designed a controlled magneto-optical inference experiment that reconstructs magnetic field, temperature, and magnetic history from a single fine-scale, high-contrast, pixel-resolved optical polarization map of feature-rich magnetic domain textures in a bismuth-substituted yttrium iron garnet film. Deep convolutional neural networks are complemented by feature-based neural-network inference using hand-crafted, physically interpretable descriptors of measured magneto-optical image data, linking the decoded information to material-dependent features and exploring their contributions. Together, these results establish magnetic texture as a high-fidelity record of external conditions enabling accurate single image multiparametric sensing and paving the way for data-driven explorations of complex magnetic states. Uncovering the physically interpretable features that encode this record sheds new light on the physics of magnetic domain formation.

## 1. Introduction

Magnetically ordered materials can host non-trivial spatial arrangements of magnetic moments known as magnetic textures in dependence of the physical magnetic material properties. In magnetic films with perpendicular magnetic anisotropy (PMA), growth-induced magnetoelastic anisotropy can establish an out-of-plane easy axis of anisotropy, while competing short-range magnetic exchange and long-range dipolar interactions give rise to the formation of magnetic domains and domain walls.[1–3] Together with stochastic domain nucleation, this may lead to the formation of typical maze domain patterns.[3–5] The magnetic domain period is determined by the relative strengths of competing interactions, which vary with parameters such as temperature and applied magnetic fields.[5, 6] Analogous to other systems with modulated phases that emerge in various physical-chemical systems, such as ferroelectric and superconducting films, as well as Turing patterns, small fluctuations in the initial conditions give rise to different realizations of otherwise similar magnetic structures.[7–10] Pattern formation is governed by the underlying energy landscape shaped by exchange, dipolar, magnetic anisotropy, and magnetoelastic energy contributions. The relative balance of these contributions is coupled to external parameters such as the applied magnetic field, the material temperature, and the magnetic history, which will influence the formation of domain patterns in a stochastic and history-dependent way. Advanced analysis methods are required to resolve how these nonlinear and history-dependent contributions shape the resulting domain patterns. Artificial neural networks (ANNs) provide such a framework by learning nonlinear mappings between high-dimensional experimental data and the physical parameters encoded in complex spatial patterns.[11–13] At their core, ANNs consist of interconnected artificial neurons, each performing a simple classification or regression operation.[11,14] In deep learning, neurons are arranged in multiple layers that transform input data into increasingly abstract representations.[11,14] The theoretical foundations of modern deep learning are closely connected to statistical physics.[15] Hopfield‘s formulation of neural networks as energy-minimizing systems, together with Hinton’s probabilistic and distributed learning framework, established a view of learning as inference in a high-dimensional energy landscape shaped by model parameters and experimental data.[15–18] Image data, such as obtained from magnetic texture analysis, is intrinsically high-dimensional, making fully connected architectures inefficient.[19] Convolutional neural networks (CNNs) address this by using shared kernels that reduce trainable parameters while exploiting spatial correlations as present in magnetic domains.[20] Pooling further lowers dimensionality and enables hierarchical feature extraction.[21] With advances such as layer-wise pretraining,[17] they have enabled high-dimensional image inference across medical,[22,23] environmental,[24] and physics-informed imaging tasks.[25]

Due to their effectiveness, CNNs have been applied to infer experimental conditions and material parameters from image data of magnetic texture. First approaches rely on fully synthetic datasets, training CNNs either on Monte-Carlo-generated spin or domain configurations to infer magnetic anisotropy, exchange, or effective magnetic fields.[26,27] Furthermore, micromagnetic simulations have been used to estimate material parameters and classify e.g. the polarity of simple magnetic ground states such as vortices.[28, 29] They are used as well to predict magnetic hysteresis properties from simulated microstructures using latent representations.[30] These studies already highlight the information density contained in complex magnetic microstructures. To bridge theory and experiment, several studies apply models trained on simulated domain textures to experimental data.[31–33] In this case, performance is limited by simulation fidelity, preprocessing steps such as binarization, and differences in imaging modality. These works represent an important first step in demonstrating that neural networks can provide access to otherwise difficult-to-measure quantities, but they do not address to what extent this information is quantitatively encoded in the domain morphology. Only a few studies train on experimental domain images. For instance, the TbCo material composition of magnetic films is inferred from magneto-optical Kerr effect microscopy data of metastable magnetic domain states including hysteresis effects.[34] The work focuses on single-target inference and shows that the domain morphology encodes the material composition through related magnetic material property variations. The analysis relied on binarized images with reduced information content. Furthermore, the applied magnetic field is not considered as a variable to be inferred. The focus is solely on the demagnetized or magnetic ground state.

Beyond classification, CNNs enable inverse inference from spatial patterns.[35,36] As a result, the underlying physical material parameters and history that generated a given magnetic texture, such as external fields or material properties, may be inferred from the observed spatial structure rather than predicting the pattern from known conditions. In particular, Schnörr et al. established a theoretical framework demonstrating that neural networks constitute a robust and statistically consistent method for inferring underlying physical parameters from complex spatial patterns, such as Turing-type structures.[36] These types of structures closely resemble the typical maze or labyrinth domain states in magnetic materials. Here, we apply this inverse-inference concept to quantitative magneto-optical images of an epitaxial Bi:YIG system with PMA. This combines Turing-type domain morphology with high magneto-optical signal-to-noise ratio and low defect density to quantify the available information about applied field, temperature, and magnetic history stored within a single magnetic domain pattern.

To demonstrate that probabilistic inference enables high-resolution, quantitative reconstruction of external parameters from magneto-optical maps of such magnetic domain patterns, artifact-free, high-contrast imaging is essential. A polarization microscope with a Stokes-polarization camera is used to obtain fully quantitative, pixel-resolved maps of the spatial magneto-optical rotation of polarization $\beta_M$.[37–39] This method reduces artefacts related to illumination, alignment, optical nonlinearities, and preprocessing. The captured domain patterns directly reflect the underlying magnetic texture, as $\beta_M$ is proportional to the out-of-plane magnetization. The quantitative approach yields superior reproducibility and physical interpretability of the data, allowing for direct application in machine learning. To maximize the dynamic range for the subsequent probabilistic inference, an epitaxial Bismuth-substituted yttrium iron garnet (Bi:YIG) film with PMA is employed, as it exhibits a large ratio between light absorption and magneto-optical Faraday rotation.[40] The epitaxial film structure results in a low defect density. The PMA is generated from a growth-induced magnetoelastic anisotropy of the YIG grown on a gallium gadolinium garnet substrate. Such high quality films have been used as bubble memory devices.[41–43] In combination with a back mirror, such films are used as magneto-optical indicator films (MOIFs) for various sensing applications.[41,44] Their primary use is to image magnetic stray fields of electrical currents,[45,46] superconductors,[47,48] or for indirect magnetic domain imaging.[49,50] Additionally, several techniques to magneto-optically measure temperature using MOIFs have been proposed. The temperature of the magnetic indicator material can be extracted either individually via changes in quantified magneto-optical contrast,[51] magneto-optical susceptibility,[52] the shape of the hysteresis loop,[53] or in parallel with the magnetic field for magneto-optical multiparametric sensing.[39,45,54,55]

## 2 Results

### 2.1 Magneto-Optical Measurements

In out-of-plane magnetic saturation, the used MOIF exhibits a large magneto-optical polarization rotation after reflection of $\beta_{M,sat} = 7.57°$ at room temperature, as can be seen in the magneto-optical hysteresis loop shown in **Figure 1**a. Upon field reversal, magnetic domains nucleate, producing equivalent but distinct periodic maze patterns with periodicity $P$, as sketched in Figure 1b. The resulting domain textures are inherently statistical in nature, as they are the result of stochastic nucleation processes.[36,54] A measured example of such a feature-rich spatial maze domain structure is shown in Figure 1c at negative coercive field $-\mu_0 H_c \approx -0.1$ mT at a temperature of 27 °C. Two states of magnetic history are distinguished by recording domain textures on either branch of the hysteresis loop. After positive saturation at +5 mT, the domain texture is recorded on the descending branch of the hysteresis loop, whereas negative saturation at –5 mT prepares the ascending branch. To investigate temperature effects on the magnetic texture the same field-history protocol was repeated at different sample temperatures. At elevated temperatures for the same applied field, the contrast decreases from Figure 1c to Figure 1d, due to a reduction of saturation magnetization $M_s$ and consequently the maximal rotation $\beta_{M,sat}$. Additional changes in the morphology in the pattern in Figure 1c and d are observed with increasing temperature. A reduction of the characteristic domain period $P$ caused by a temperature-induced shift in the balance between magneto-static and domain wall energy contributions is seen.[4,56] Moreover, the favored alignment of the domains along three in-plane directions (120° apart), originating from the distinct cubic magnetocrystalline anisotropy with easy axis along the ⟨111⟩ directions in the epitaxial (111) film, seems more pronounced for higher temperatures. These morphological differences in magnetic domain configurations provide information that can be exploited in inference experiments. Standard magneto-optical magnetometry measurements record the signal without spatial information, which inherently results in significant loss of information. From an inference perspective, this presents a fundamental limitation of an uncertainty up to the coercivity field $H_c$. This ambiguity can be reduced by taking full spatial information into account. Although the magnetic domain structures of the descending branch at $-\mu_0 H_c$ (Figure 1d) and of the ascending branch at $+\mu_0 H_c$ (Figure 1e) have the same average magnetization, they exhibit inverted magnetic domain pattern characteristics. Consequently, morphological features change depending on the branch of the hysteresis loop and the magnetic history is encoded in the magnetic texture. One such characteristic feature is the number of endpoints of magnetic domains of a certain magnetization. Here, endpoints are defined as terminal points of the medial axis of band magnetic domains, as shown in Figure 1b. These terminations form pairs with branching points in domains of opposite magnetization, which may be spatially separated. In particular, the number of endpoints of domains with positive magneto-optical rotation $N_{+,ends}$ is more than five times higher in Figure 1d than in Figure 1e, while the domain period is quasi constant. The nearly constant domain period suggests that the characteristic length scale is determined by the temperature, while the endpoint density captures the stochastic, history-dependent topology of domain growth and nucleation. A total of 11,372 quantitative polarization maps were acquired within a magnetic field range between −1 mT and +1 mT and a temperature range between 21°C and 72°C.

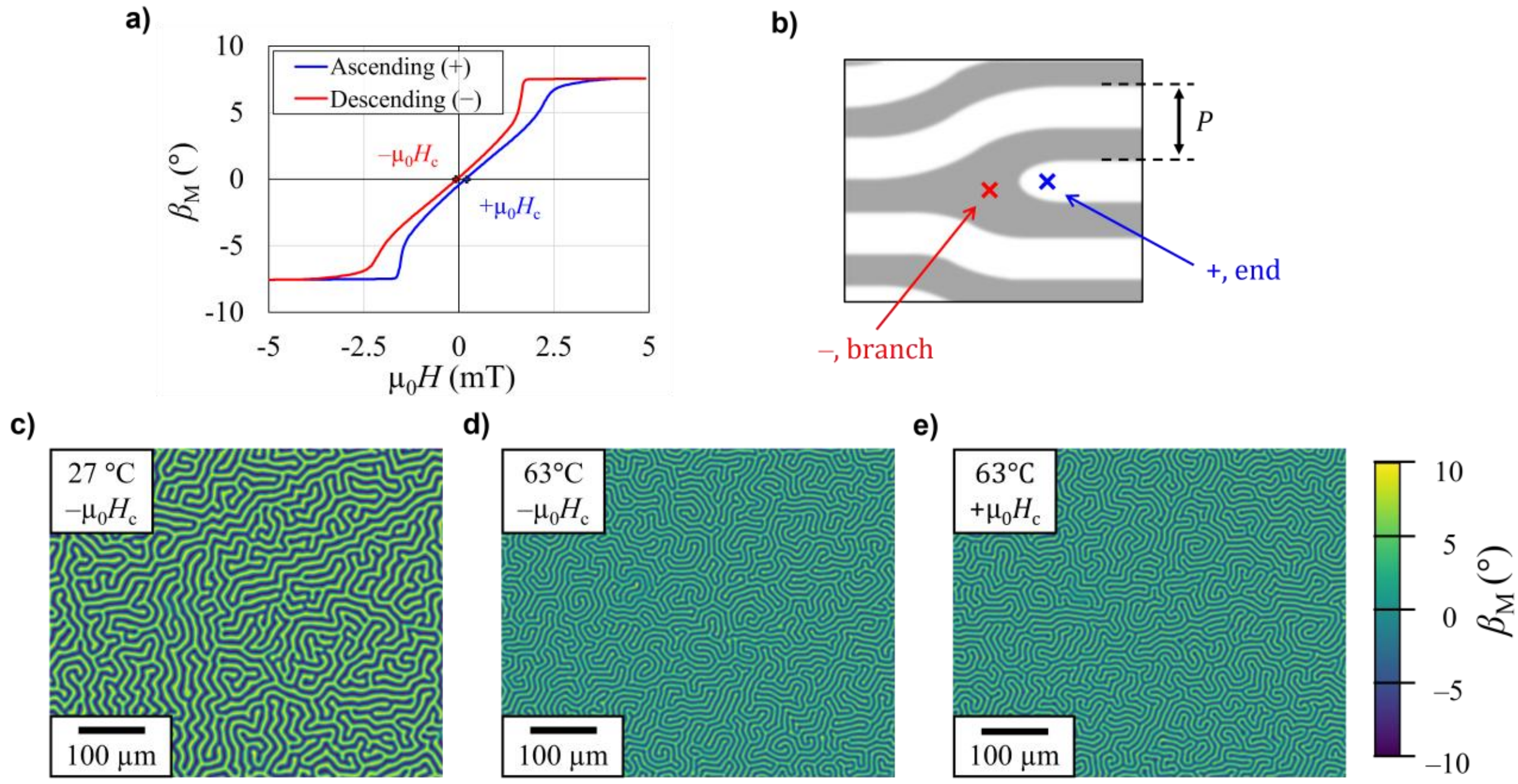

**Figure 1. Magnetic hysteresis and domain pattern.** (a) Magneto-optical rotation $\beta_M$ to applied out-of-plane field $\mu_0 H$ showing the magnetization loop of the Bi:YIG sample at 25°C. The history is differentiated by the ascending (+) and descending (−) branch of the hysteresis. (b) Schematic domain structure with indicated domain period $P$, an endpoint of a domain with positive magnetization (+, end) and a branching point of a domain with negative magnetization (−, branch). (c) Quantitative magneto-optical polarization map at negative coercivity $-\mu_0 H_c = -0.1$ mT at 27°C. Magneto-optical polarization maps at 63°C at a negative applied field of (d) $-\mu_0 H_c = -0.1$ mT and (e) of a positive applied field of $+\mu_0 H_c = +0.1$ mT for the descending, respectively ascending loop branch.

### 2.2 Inference Using Hand-Crafted Features

The underlying information of applied field and temperature in the pattern is encoded through complex non-linear dependencies. For CNNs, this information is decoded implicitly from the full image data or map such that their learned representations remain largely opaque to physical interpretation. To uncover the physical origin of the inferred information, we therefore first introduce a feature-based analysis. A set of hand-crafted physically interpretable descriptors is extracted from each polarization map, simplifying the high-dimensional and complex matrix into an interpretable 20-dimensional vector of global features. This feature vector (FV) combines first order statistics of the magneto-optical rotation $\beta_M$, such as the mean $\overline{\beta_M}$ and standard deviation $\sqrt{\langle \beta_M^2 \rangle}$, with descriptors derived from Fourier analysis, such as the characteristic domain period $P$ (including its mean value and variation), and morphology descriptors, such as the number of domains of positive magneto-optical rotation and the number of their endpoints. This complementary approach allows us to conclusively identify which aspects of spatial and statistical properties of the magnetic texture encode information about the external parameters. To perform the inference, the resulting FV is used as input to a regression model that maps these descriptors to the external parameters. The history state was treated as a regression variable during training and discretized into binary (+/−) in post-processing. Specifically, the 20 preprocessed scalar features are normalized and fed into a fully connected multilayer perceptron with 9 layers, which predicts DC magnetic field magnitude, magnetic history, and temperature.

**Figure 2** summarizes four of the most relevant extracted features and their measured dependence on magnetic field, temperature, and magnetic history across the full acquired dataset as used for the inference. Detailed information about the other features is available in the supporting information. The spatially averaged magneto-optical rotation $\overline{\beta_M}$ (Figure 2a and e) reproduces the magnetization loop, with decreasing magneto-optical susceptibility at elevated temperatures. The two history branches are separated by the nearly temperature-independent coercivity, but the average contains no information on the local domain configuration. Nevertheless, its low noise level makes it an essential contribution for accurate field inference for known temperature and history.

The standard deviation $\sqrt{\langle \beta_M^2 \rangle}$ (Figure 2b and f) captures magneto-optical domain contrast and scales strongly with temperature, reflecting the reduction of $\beta_{M,sat}$ and thus saturation magnetization, while showing only weak dependence on field and history. Like the mean, it is highly repeatable between equivalent patterns, though it is dependent on the imaging system. A complementary temperature-sensitive feature is the characteristic dominant domain period $P$, extracted from Fourier analysis (Figure 2c and g). As the $P$ decreases with temperature, the domain wall energy, dependent on exchange, growth-induced uniaxial perpendicular magnetoelastic anisotropy and magnetocrystalline anisotropy, decreases faster than the demagnetization energy set by the saturation magnetization.[4,56]

With the hysteresis loop shown in Figure 1a, a given magneto-optical rotation $\overline{\beta_M}$ (and thus average magnetization) generally can correspond to all magnetic field states within the loop area. Here, two history states are differentiated, depending on whether the field is approached from positive or negative saturation along the ascending (+) or descending (−) branch of the hysteresis loop. Such ambiguity is a well-known challenge for parameter inference in hysteretic systems, where identical macroscopic observables may correspond to different underlying states depending on the system's prior evolution. While the spatially averaged rotation $\overline{\beta_M}$ alone cannot distinguish these two states, the magnetic history is encoded unambiguously in the domain morphology. This is exemplified by the number of endpoints of positive domains $N_{+,\mathrm{ends}}$ (Figure 2d and h), which clearly differentiates the two branches of magnetic history. For the descending branch (−), a few individual negative stripes form from a small number of nucleated negatively magnetized domains within the positively magnetized matrix. During domain growth, these negative stripes start to twirl, thus forming many positive domain ends. For the ascending branch (+), the positive domain ends still originate primarily from nucleation and are thus less numerous. The behavior with magnetic field is inverted for the number of negative domain ends, thus mitigating the otherwise occurring ambiguity in the overlapping region in Figure 2d. The number of endpoints is equivalent to the number of domain branches, and therefore indicates the number of imperfections in the fundamentally striped pattern.[3,57] The number of maze domain imperfections, indicated by $N_{+,\mathrm{ends}}$ for the descending history (−), increases with temperature due to the decreasing periodicity. However, if the number of endpoints is normalized to the number of domain period area squares $P^2$ within the field of view, a decrease of imperfections with temperature of more than a factor of 2 is observed. This indicates an increased crystallinity, or decreased entropy, of the domain structure with temperature. This correlates with the increased perceived threefold symmetry for higher temperatures in Figure 1. This corresponds to a more favored alignment of the stripes at higher temperatures along the three directions defined by the cubic magnetocrystalline anisotropy, as seen in the comparison between Figure 1c to Figure 1d and e.

Both the domain period and the number of domain ends contain information about the spatial distribution of the magnetic pattern, which forms from the inherently statistical nucleation process and thus are noisy in comparison to the first order statistics.

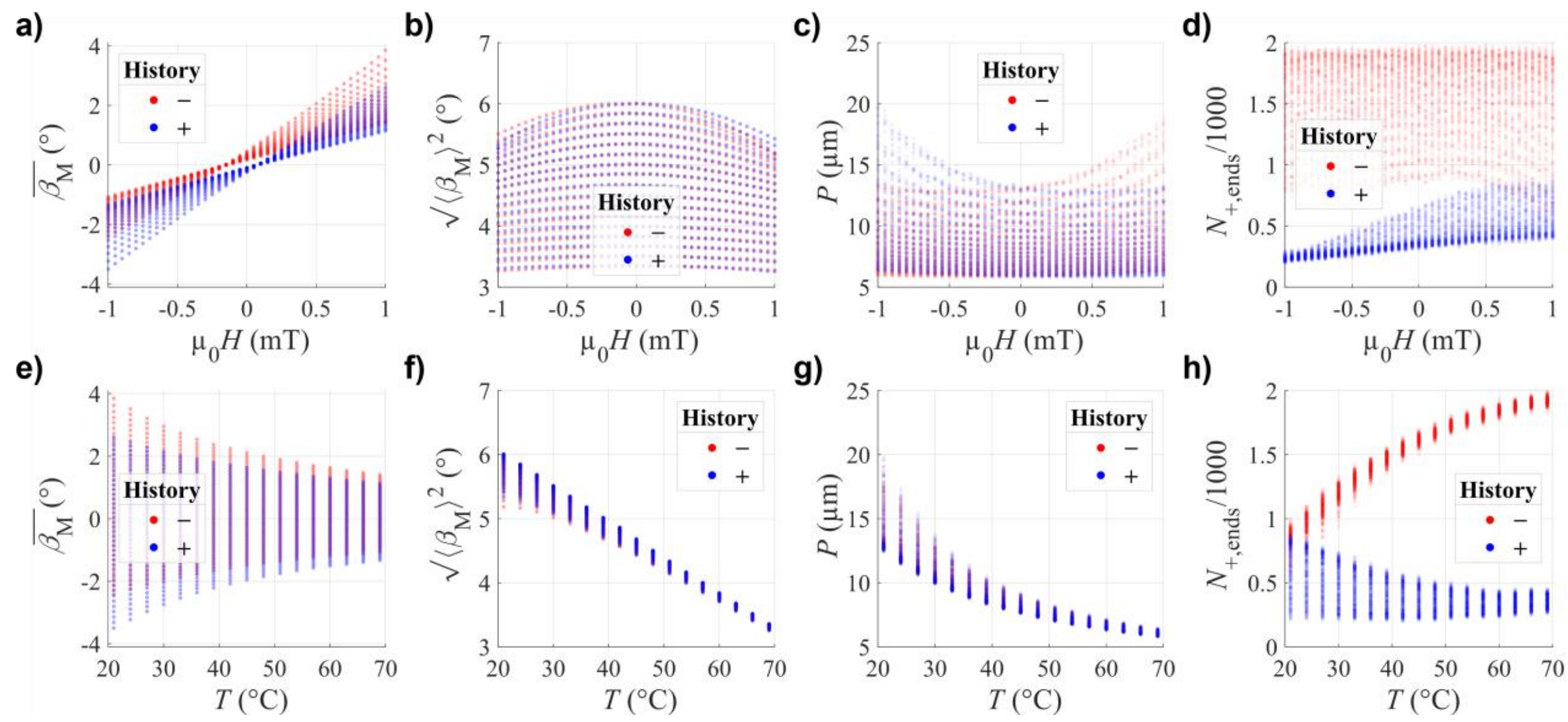

**Figure 2. Selected feature vector.** Exemplary extracted features from quantitative polarization maps to either (a)-(d) applied magnetic field or to (e)-(h) set temperature. (a),(e) Mean magneto-optical rotation $\overline{\beta_{\mathrm{M}}}$ proportional to the *z*-component of magnetization. (b),(f) Standard deviation or contrast within the map $\sqrt{\langle\beta_{\mathrm{M}}\rangle^2}$ of the magneto-optical rotation. (c),(g) Dominant domain period $P$ of the magnetic maze domains. (d),(h) Number of morphological endpoints of the positive domains $N_{+,\mathrm{ends}}$. The plots contain 11,372 data points and are differentiated between ascending branch (+) and descending branch (−) history of the full hysteresis, respectively.

### 2.3 Inference Using Convolutional Neural Networks

Since the FV uses a compact set of physically interpretable descriptors, it is inherently limited to information captured by these predefined quantities. To assess the magnitude of additional information contained in the magnetic domain texture, we next perform an inference experiment directly on the captured magneto-optical rotation maps using a CNN. By learning directly from the full spatial information, the CNN does not rely on predefined descriptors and can exploit fine-scale correlations and higher-order spatial structures that are inaccessible to global descriptors. Thus, a higher sensitivity compared to the FV is expected for the CNN.

The network consists of seven convolutional layers with increasing channel depth, each combined with group normalization and max-pooling to progressively extract hierarchical spatial features. These layers are followed by adaptive average pooling and two fully connected layers that map the learned representation onto a three-component continuous output. The history state was again treated as a regression variable during training. All inputs are normalized and separated into a training and validation dataset. To conserve GPU memory, each map is subdivided into smaller tiles before being processed by the network. Further data augmentation is implemented by vertical and horizontal flipping of the tiles to generate in total 64 sub-maps used as input for both training and validation. This procedure creates additional similar but distinct patterns, which decreases the number of necessary experimental images or maps for training and increases the statistics for inference. Additionally, the flipping breaks the three-fold symmetry and changes positions of magnetic defects in between the maps for possibly better generalization. The CNN regresses to the external parameters from the normalized, augmented tiles, and the results of all tiles are averaged for the prediction of the original map.

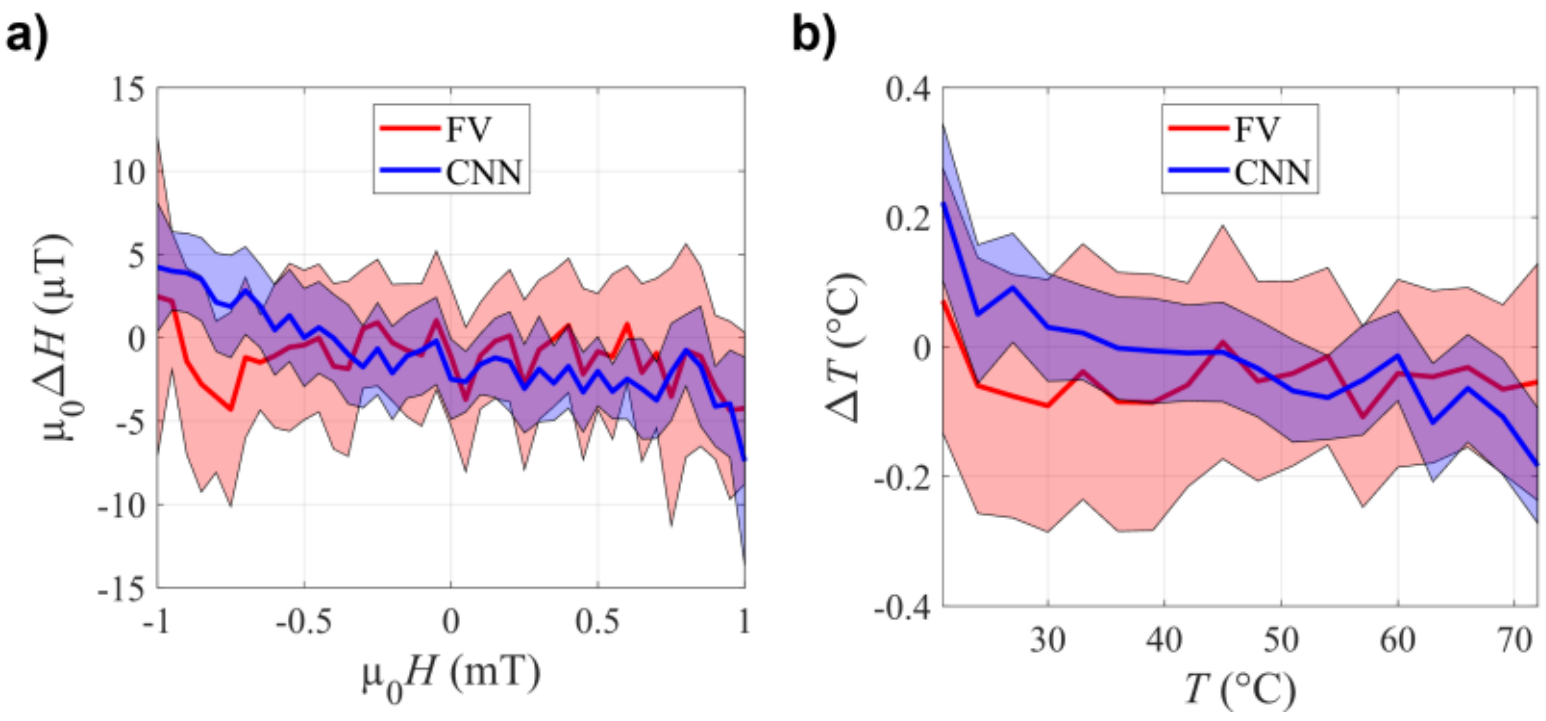

**Figure 3. Inference residuals to physical parameters.** (a) Residual magnetic field prediction $\mu_0\Delta H$ as a function of the applied field $\mu_0 H$ for the machine learning models based on the global feature vector (FV) and on the full spatial magneto-optical map using a convolutional neural network (CNN). Solid lines represent the mean residual (systematic bias) while shaded regions indicate the root-mean-square accuracy. (b) Equivalent residual temperature prediction $\Delta T$ plot over the applied temperature $T$.

The measured dataset is randomly divided into training and validation subsets (90% | 10%). To evaluate the performance of the CNN-based inference in comparison to the FV approach, both trained models are applied to the same independent validation dataset. **Figure 3** compares the inference performance of both approaches for reconstructing the applied magnetic field $\mu_0 H$ and temperature $T$. Figure 3a and b show the residuals $\mu_0\Delta H$ and $\Delta T$ as a function of the true applied parameters. The mean residual indicates systematic bias, while the statistical root-mean-square spread of the predictions represents the extracted information content from the polarization maps.

For both, $\mu_0 H$ and $T$, the FV-based inference yields residuals centered close to zero over the full parameter range, and thus exhibits negligible systematic bias. The statistical spread reflects the intrinsic information content captured by the extracted features. Quantitatively, the FV approach achieves average sensitivities per map of $\mu_0\Delta H_{FV} \approx 6.5$ μT (0.33% FS, full scale) and $\Delta T_{FV} \approx 0.17$ K (0.33% FS), while correctly identifying all history states. The CNN further improves the inference accuracy by exploiting the full spatial information of the magnetic pattern, resulting in enhanced sensitivities of $\mu_0 \Delta H_{CNN} \approx 3.8\ \mu T$ (0.19% FS) and $\Delta T_{CNN} \approx 0.12$ K (0.24% FS), without loss of accuracy in the inferred history states. The weak residual trend with both magnetic field and temperature indicates a remaining systematic error that we attribute to slight underfitting of the regression model. As the magnitude of this trend is smaller than the statistical uncertainty, its impact on the reported sensitivities is negligible. Neither approach shows a significant dependence of the statistical spread on the external parameters within the explored range.

To investigate the dependence of inference performance on dataset size, both models are evaluated using varying numbers of training maps. The resulting evolution of the average sensitivities is shown in **Figure 4**. Next to the root mean square residual of magnetic field (Figure 4a) and temperature (Figure 4b), the regression variable for the history state $y_{hist}$ is described as well (Figure 4c). The binary values are defined as either +1 or –1, thus a residual of at least 1 corresponds to a falsely identified history state.

As expected, the FV approach outperforms the CNN for small training datasets, consistent with reduced data requirements at lower model complexity.[36] Remarkably, both approaches already achieve high accuracy with very limited data. Using only 57 training maps, the FV method yields sensitivities of approximately 90 μT (4.5% FS) for the magnetic field and 1 K (2% FS) for temperature, while correctly identifying the magnetic history in all cases. With increasing dataset size, the performance of both models improves and eventually saturates. The FV approach reaches its performance plateau earlier, at approximately 4,700 training maps, but at a higher residual level compared to the CNN, which continues to improve up to around 5,700 maps. The observed peak for the FV approach is part of relatively large fluctuations between different trainings. It indicates a higher sensitivity to initialization and training conditions, reflecting a reduced robustness of the FV approach compared to the CNN.

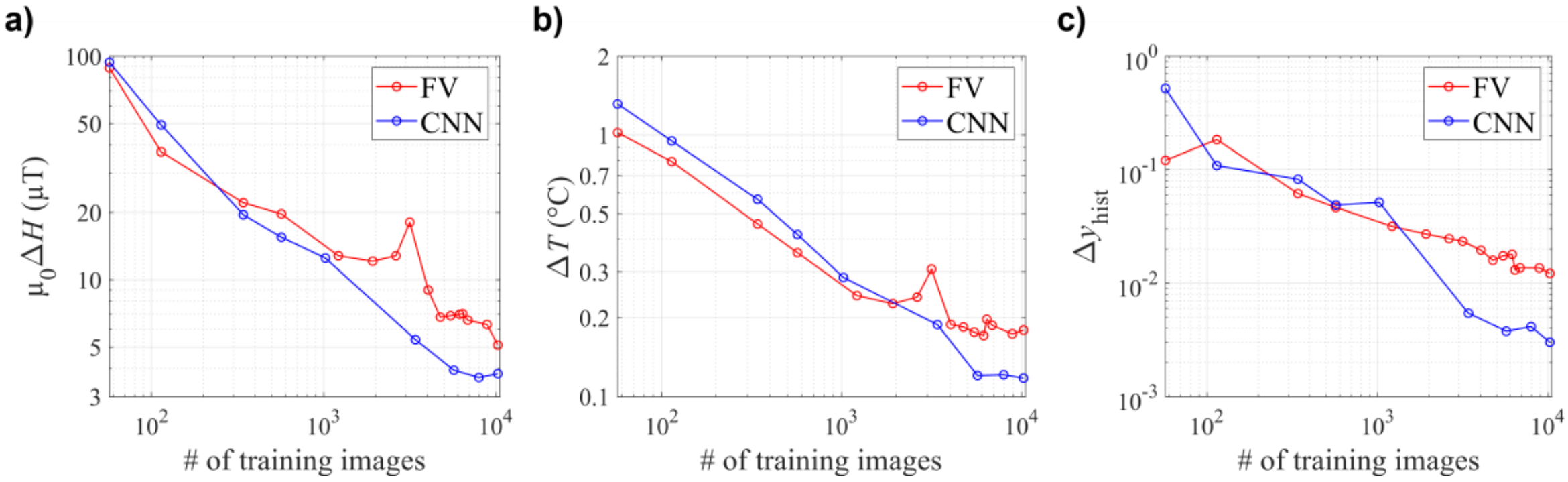

**Figure 4. Inference residuals to training data.** (a) Average residual magnetic field prediction $\Delta\mu_0 H$ to number of used training maps based on the global FV and using full maps (CNN). (b) Equivalent residual plot for the temperature $T$. (c) Average residual of the history regression parameter $\Delta y_{hist}$ in percent to the number of training maps. A residual of at least 1 of a single datapoint corresponds to a falsely identified history state.

## 3. Discussion

These results demonstrate that a single quantitative magneto-optical image of the magnetic texture contains sufficient information to simultaneously reconstruct both past and present magnetic field and temperature with high precision. This highlights the remarkably high information density encoded in magnetic domain configurations, which act as a fingerprint of the underlying energy landscape and reflect the combined influence of applied field, temperature and magnetic history, including the branch of the hysteresis loop.

A key advantage of the CNN over the FV approach lies in its ability to exploit fine-scale morphological details and higher-order correlations that are difficult to identify and quantify using manually defined features. The improved performance of the CNN therefore reflects its capacity to access a broader fraction of the information content inherent in the magnetic pattern. Nonetheless, as the average residuals between CNN and FV differ by less than a factor of 2, a large share of the information content is successfully identified, extracted and interpreted using the computationally less expensive global features. The only slightly reduced inferential performance of the FV is consistent with the findings in ref. [34].

The achieved temperature sensitivity can be related to the underlying magnetic parameters. The relative magneto-optical rotation in saturation $\beta_{M,sat}$, which in close approximation scales with the saturation magnetization $\boldsymbol{M}_s$, decreases by approximately 20% between 20 °C and 60 °C, corresponding to a rate of about 0.5% K$^{-1}$. The achieved CNN accuracy of 0.12 K therefore corresponds to a resolution of relative changes in $\boldsymbol{M}_s$ on the order of 0.06% for a single image. Given sufficiently random domain nucleation, many such images could be used for a stochastic analysis. Since multiple magnetic parameters depend on temperature, this sensitivity is effectively distributed across the full set of temperature-dependent quantities. Future experiments could aim to disentangle also these contributions. Additionally, as the minimal residual of the history state is an order of magnitude lower than necessary for faultless decoding, a more fine-grained identification of the magnetic and temperature history could be deduced as well.

Compared to existing machine-learning-based approaches analyzing magnetic domain textures, the present method systematically explores the boundaries of the information contained within experimental data dependent on external control parameters, rather than aiming to derive specific magnetic parameters. The field of view is chosen such that the magnetic domains are consistently resolved while maximizing the observed area. The epitaxial YIG film provides a fine, regular and highly variable domain structure with a periodicity much smaller than the distances between defects. This allows Turing type patterns to form which are highly susceptible to small changes in the governing parameters and initial conditions.[58] Combined with the high magneto-optical signal-to-noise ratio of the MOIF, these properties maximize the information content within the fingerprint domain patterns. Furthermore, as different physical quantities are addressed in each existing approach, a direct comparison of existing machine-learning approaches to the presented method is inherently challenging and quantitative comparisons of the relative errors of the underlying magnetic parameters should be interpreted carefully. The present method represents an improvement of the achievable relative accuracy to existing approaches of more than one order of magnitude.[28,31–33] This emphasizes the importance of a large, quantitative high quality training dataset. Importantly, the inclusion of magnetic history in the inference of the present method directly addresses the multivariate and partially ambiguous mapping between the morphology of magnetic textures and external parameters arising from magnetic hysteresis.

From an application perspective, the achieved sensitivities can be expressed in a time-normalized form. Assuming a measurement cycle consisting of image acquisition and domain reset (0.25 s in the present setup) and white-noise statistics, this yields normalized sensitivities of $\Delta T_{\mathrm{CNN,norm}} \approx 0.06$ K Hz $^{-1/2}$ and $\mu_0\Delta H_{\mathrm{CNN,norm}} \approx 1.9$ µT Hz $^{-1/2}$ for the CNN, compared to $\Delta T_{\mathrm{FV,norm}} \approx 0.09$ K Hz $^{-1/2}$ and $\mu_0\Delta H_{\mathrm{FV,norm}} \approx 3.3$ µT Hz $^{-1/2}$ for the FV approach. Given the high signal-to-noise ratio of the magneto-optical images, further improvements are readily achievable by reducing the exposure time. In combination with fast magnetic pulsing, the ultimate limitation is expected to be set by the camera frame rate. Differences in measurement frequency readout, used sensor material and area limit the comparability to existing magneto-optical sensing methods. Single parameter readouts with a similar sample achieve accuracies that are more precise for both temperature and magnetic field than the present method.[51,52] In contrast, examples of multiparametric readout are rare and exhibit comparable accuracies, partly even with spatial resolution.[45,55] The additional inference of a past magnetic field history in combination with temperature has applications beyond simple sensing applications.

## 4. Conclusion

In summary, the results quantify the information contained about the underlying energy landscape by the magnetic texture of a near optimal experimental system using a perpendicular magnetic anisotropy YIG with high magneto-optical contrast. The presented approach uniquely enables simultaneous, multiparametric inference of magnetic field and temperature as external parameters and extraction of memory of magnetic treatment from a single magnetic domain image of a magnetic film. The boundaries of the methodology are investigated by integrating magnetic domain behavior with sophisticated analysis. Expanding the demonstrated framework to include additional quantities could expand sensing capabilities and provide access to additional correlations between magnetic domain patterns and the underlying magnetic parameters. This opens a route to further study the decoding of complex spatial structures of the magnetic energy landscape for other magnetic systems and quantities. Future work may therefore focus on improving generalization by training across physical quantities, material systems, samples, defect distributions and dynamic states, thereby advancing magneto-optical imaging toward a versatile platform for fast multiparametric sensing and increased data-driven exploration of complex magnetic states.

## 5. Methods

*Magnetic Material:* The magnetic sample is a magneto-optical indicator film (MOIF). It consists of a bismuth-substituted (semi-)transparent ferrimagnetic yttrium iron garnet (Bi:YIG) film epitaxially grown on a (111)-oriented gallium gadolinium garnet (GGG) substrate, exhibiting growth-induced perpendicular magnetic anisotropy.[44] The film has a thickness of approximately 3 µm and a Néel temperature of 138 °C. A mirror and a protective layer are deposited on the magnetic film side. The maximal measured magneto-optical polarization rotation after reflection during out-of-plane saturation is $\beta_{\mathrm{M,sat}} = 7.57°$ for white light at $T = 25$ °C. The MOIF sensor is commercially available as “Type A” from matesy GmbH.[59]

*Parameter Control:* The magnetic field is applied with an out-of-plane coil in Helmholtz configuration driven by a four-quadrant voltage controlled current source. All measurements are carried out in a quasi-DC state. For temperature control, the setup is water-cooled with active PID control of an ohmic heater situated beneath the sample. The temperature is measured via a thermistor within the sample holder. The sample is thermally connected to a copper plate with thermal paste, which in turn is mechanically screwed into physical contact with the heating unit. The long-term temperature stability of the sample is measured magneto-optically using the method based on the work by Kustov et al. with a long-time root mean square deviation of 0.014 °C.[51]

*Imaging System:* The experimental sensor setup is based on a magneto-optical wide-field microscope with a white light source with Köhler type illumination in reflection.[37] All measurements are performed using a 5x objective with a numerical aperture of 0.16, resulting in a field of view of 916 µm × 1096 µm. The total exposure time per image is 106.7 ms. The illumination aperture in the back focal plane is centered to ensure pure polar magneto-optical sensitivity. To ensure measurement comparability, the microscope focus is readjusted before each measurement at a new temperature step on the mirror-YIG interface by maximizing the image standard deviation, compensating for thermal expansion of the sample holder.

The magneto-optical rotation maps are acquired using a Stokes-polarization camera using a Sony CMOS Pregius Polarsens IMX250MZR Sensor chip with a 3.45 µm pixel size.[38,39,60,61] This enables quantification of the angle of linear polarization, which scales linearly with the Faraday rotation and, consequently, with the magnetization perpendicular to the film plane. The quantitative imaging scheme provides invariance in light source intensity, illumination gradients, and defects in the optical path.

*Computing:* Training and inference based on the FV method was carried out by minimizing the mean-squared error using the Adam optimizer. Calculations were performed on an Intel® Xeon® CPU E5-1630 v4 @ 3.7 GHz. The (non-optimized) calculation of the feature vector from a single image takes 16.2 s. Training with the full dataset of the neural network takes 19.4 minutes with 400 epochs. Inference per image takes 90 µs. Training for the CNN is carried out by minimizing the mean-squared error using the Adam optimizer. In contrast to the FV model, the CNN is computationally substantially more demanding. All CNN training and inference were carried out on a NVIDIA H100 GPU equipped with 18,432 CUDA cores, 80 GB of GPU memory, and 2 TB/s memory bandwidth. Training with the full dataset takes 2,320 minutes with 70 epochs. Inference per augmented image takes 0.138 s. See supporting information for details on the architecture of both models.

**Acknowledgements**

Funded by the Deutsche Forschungsgemeinschaft (DFG, German Research Foundation) MC 9/20-2 (Project No. 395252182) and VO 3017/1-1 (Project No. 571878852). M.V. thanks the priority research area KiNSIS (Kiel Nano, Surface and Interface Science) at Kiel University for their support through the KiNSIS Early Career Grant.

**Data Availability Statement**

The datasets generated and analyzed during the current study are available from the corresponding author on reasonable request.

**Decoding Magnetic Texture – Supporting Information**

*Michael P. Path*, Jeffrey McCord, Michael Vogel**

Michael P. Path, Jeffrey McCord, Michael Vogel

Nanoscale Magnetic Materials - Magnetic Domains, Department of Materials Science, Kiel University, Kiel, Germany

E-mail: mipa@tf.uni-kiel.de, mv@tf.uni-kiel.de

Jeffrey McCord, Michael Vogel

Kiel Nano, Surface and Interface Science (KiNSIS), Kiel University, Kiel, Germany

**Features**

All 20 measured features are calculated from the measured magneto-optical rotation maps $\beta_{\mathrm{M}}$ measured at via the Stokes polarization camera. The images are processed using Python. The features are listed below, together with their behavior to past and present magnetic field and temperature, as well as a short explanation.

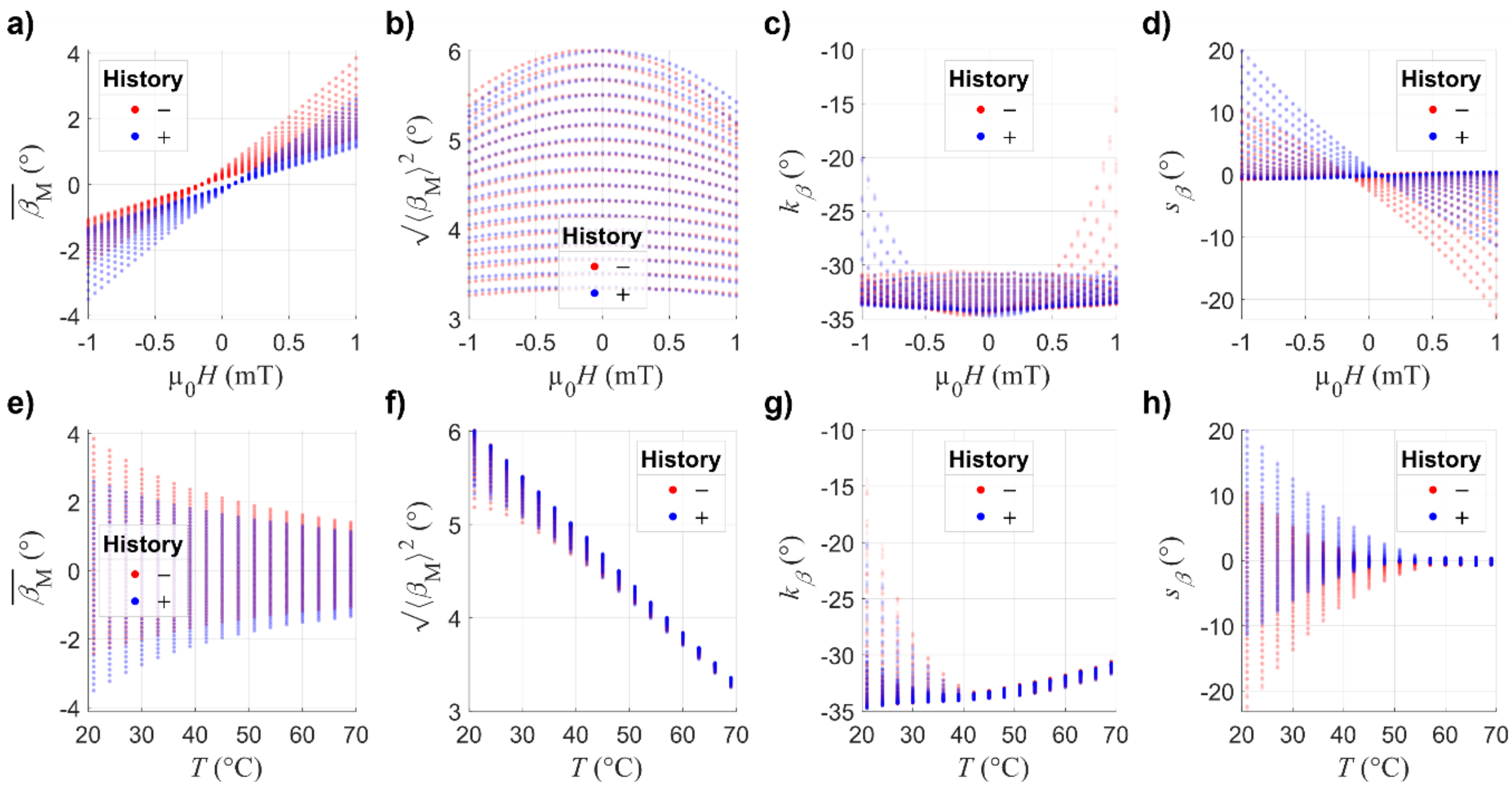

**Figure S1.**

$\overline{\beta_M} = \frac{1}{n}\sum_{i=1}^{n} \beta_{\mathrm{M}}$

**Fig. S1a,e:** The mean of magneto-optical signal in ° indicates the average perpendicular magnetization $\beta_{\mathrm{M}} \sim M_{\mathrm{z}}$.

$$\sqrt{\langle\beta_M\rangle^2} = \sqrt{\frac{1}{n}\sum_{i=1}^{n}\left(\beta_{\rm M} - \overline{\beta_M}\right)^2}$$

**Fig. S1b,f:** The standard deviation of the distribution of the magneto-optical rotation map in ° is an indicator for the contrast of the map and thus scales primarily with the saturation magnetization. It is dependent on the magnification and focus.

$$k_\beta = \frac{1}{n}\sum_{i=1}^{n}\left(\frac{\beta_M - \overline{\beta_M}}{\langle\beta_M\rangle^2}\right)^4$$

**Fig. S1c,g:** The Kurtosis is the fourth standardized normalization of the of the distribution of the magneto-optical rotation map. It is an indicator for the asymmetry of the distribution of $\beta_{\rm M}$.

$$s_\beta = \frac{1}{n}\sum_{i=1}^{n}\left(\frac{\beta_M - \overline{\beta_M}}{\langle\beta_M\rangle^2}\right)^3$$

**Fig. S1d,h:** The skewness is the third standardized normalization of the distribution of the magneto-optical rotation map. It is an indicator for the probability of values close to the edges of the distribution of $\beta_{\rm M}$.

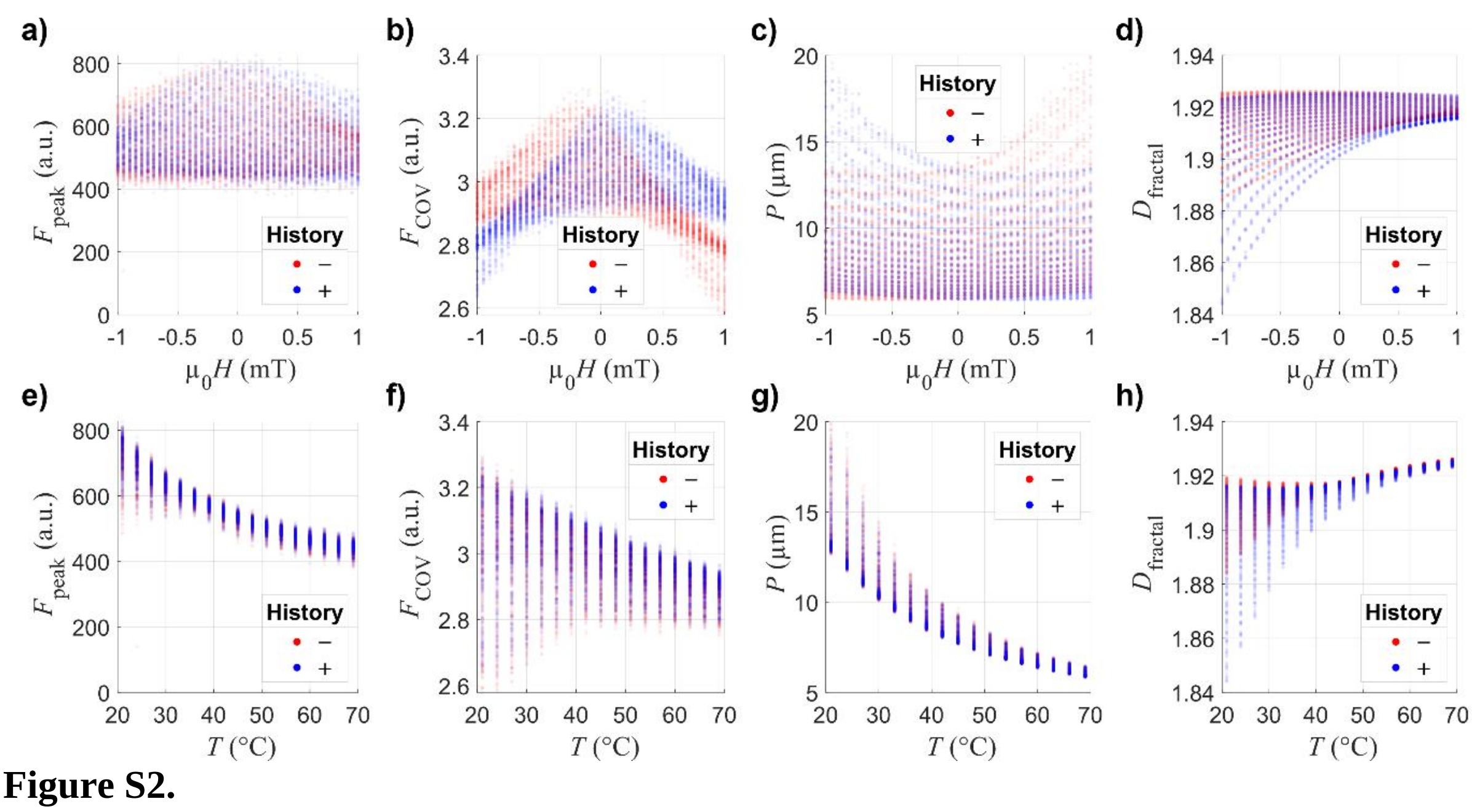

**Figure S2.**

The radial profile $F$ of the Fourier transform of the $\beta_{\rm M}$ rotation map is calculated.

$F_{\rm peak}$ — **Fig. S2a,e:** Peak height of the radial profile caused by the primary domain period. A higher value indicates a larger saturation rotation and a sharper distribution of the domain frequencies.

$F_{\rm COV}$ — **Fig. S2b,f:** Coefficient of variation of the radial profile. It is an indicator for frequencies within the domain pattern besides the primary one.

$P$ — **Fig. S2c,g:** Domain period calculated from the frequency with the largest peak in the radial profile. It scales with the temperature dependent balance of domain wall energy and demagnetization energy.

$D_{\text{fractal}}$ **Fig. S2d,h:** Fractal dimension of the positive domains within the image and is a scale for the complexity of the pattern. It increases with temperature due to the reduced domain period and is lowest close to saturation (so at negative field after negative saturation), as less twirling could have happened during domain growth.

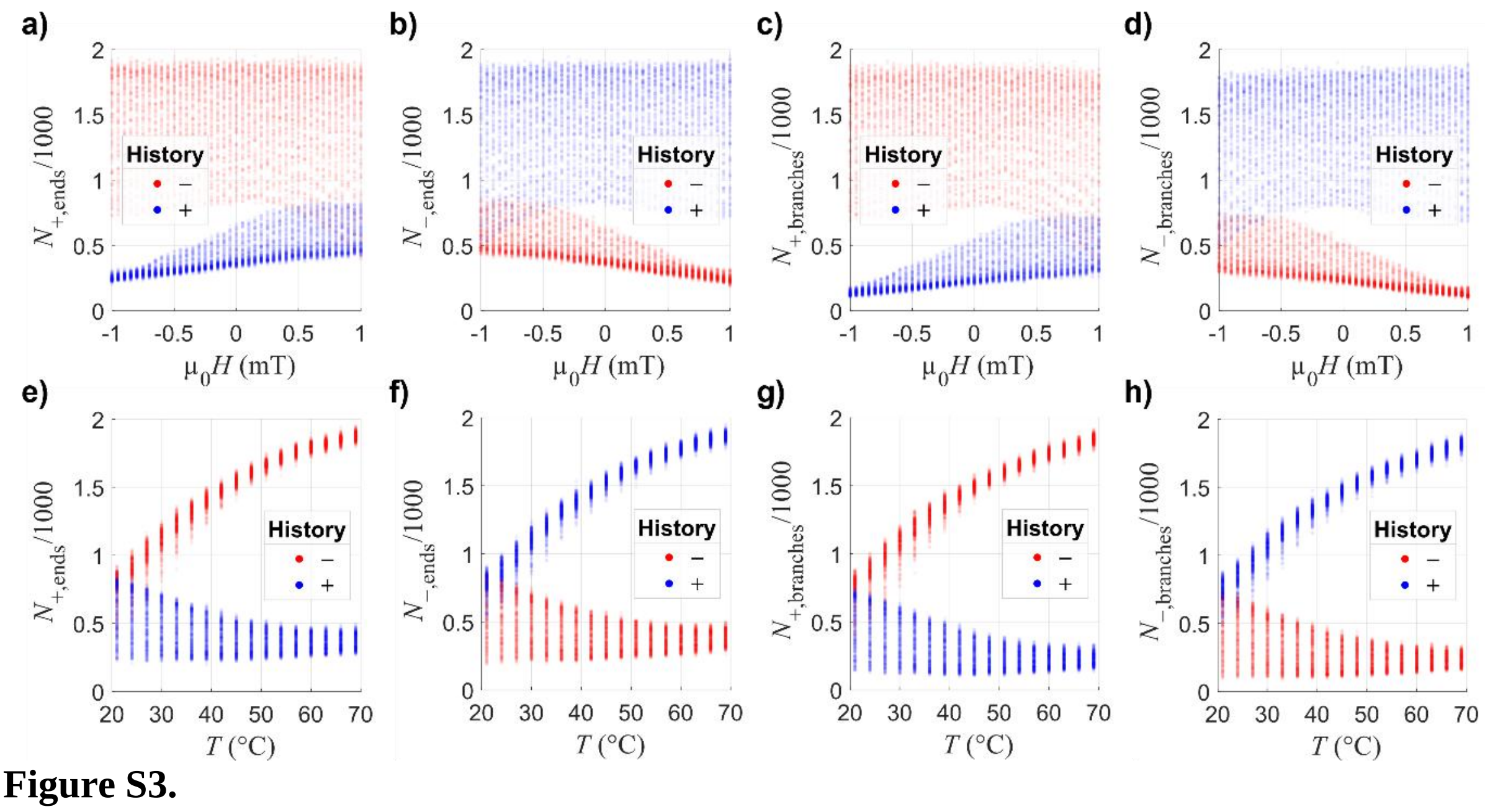

**Figure S3.**

The image is binarized with either positive (+) or negative (–) domains as positive and then skeletonized, so the domains are reduced to a width of 1 pixel. For each pixel part of the skeleton, the number of neighbors also being part of the skeleton are counted. If it is one neighbor, it is counted as an end, if it has 3 or more neighbors, it is counted as a branch.

$N_{+,\text{ends}}$ **Fig. S3a,e:** Total number of endpoints of the positive domains.

$N_{-,\text{ends}}$ **Fig. S3b,f:** Total number of endpoints of the negative domains.

$N_{+,\text{branches}}$ **Fig. S3c,g:** Total number of branches of the positive domains.

$N_{-,\text{branches}}$ **Fig. S3d,h:** Total number of branches of the negative domains.

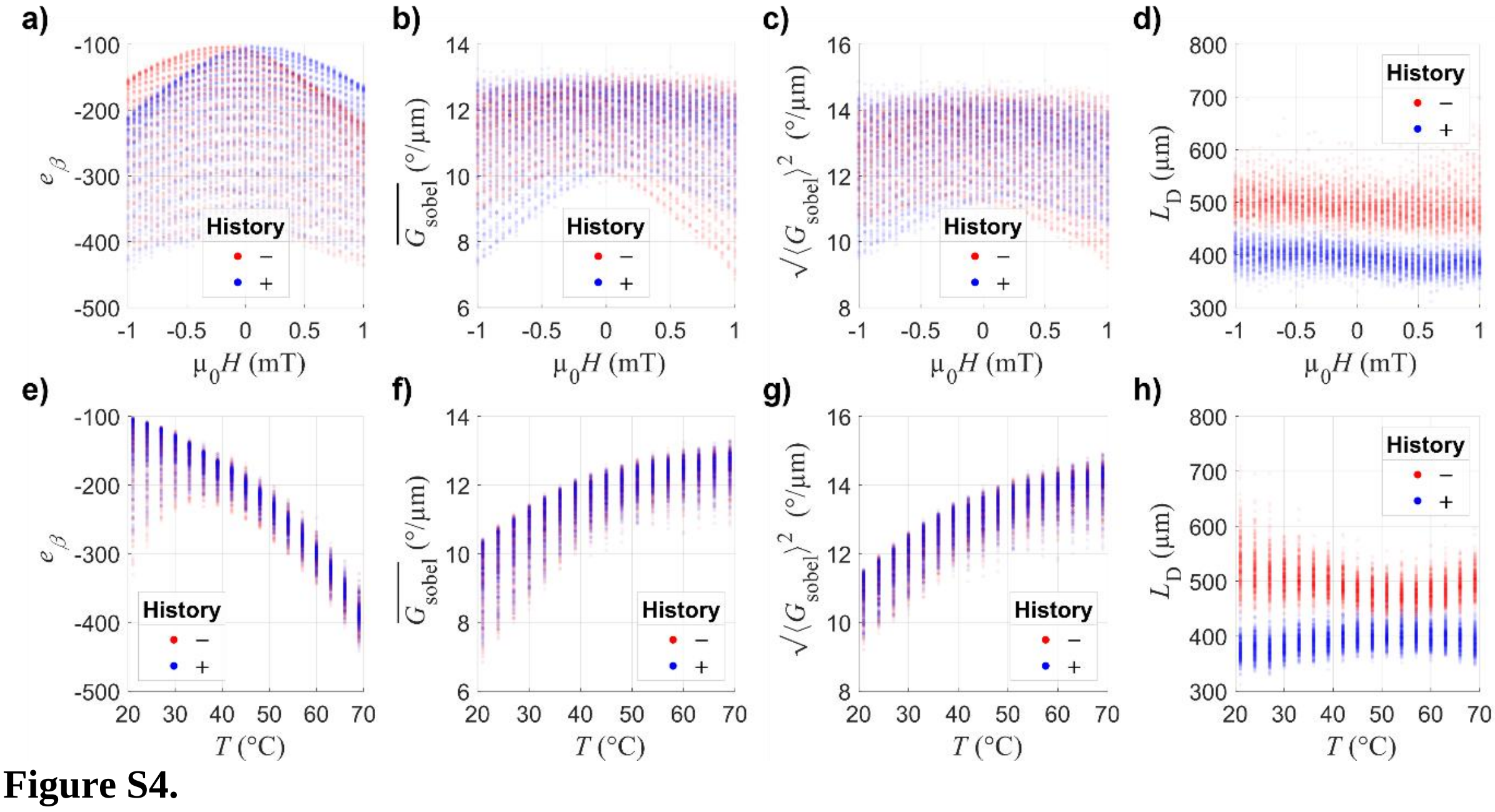

**Figure S4.**

$e_\beta$ — **Fig. S4a,e:** Parameter calculated similarly to image entropy, but probability density instead of probability has been used, causing negative values. No impact on information content is expected.

$\overline{G_{sobel}}$ — **Fig. S4b,f:** Average gradient within the image using calculated the Sobel operator. It increases with temperature due to the increasing domain wall density despite a lowered gradient within the domain walls due to the lowered saturation rotation. It decreases with the magnitude of applied field, as the domain wall density decreases.

$\sqrt{\langle G_{sobel}\rangle^2}$ — **Fig. S4c,g:** Standard deviation of gradient within the image calculated using the Sobel operator.

$L_D$ — **Fig. S4d,h:** Average total length of connected positive magnetic domains calculated from total pixel count of domains in the skeletonized image. Positive domains tend to be longer in the descending history state coinciding with the number of domain ends, though without the temperature dependence.

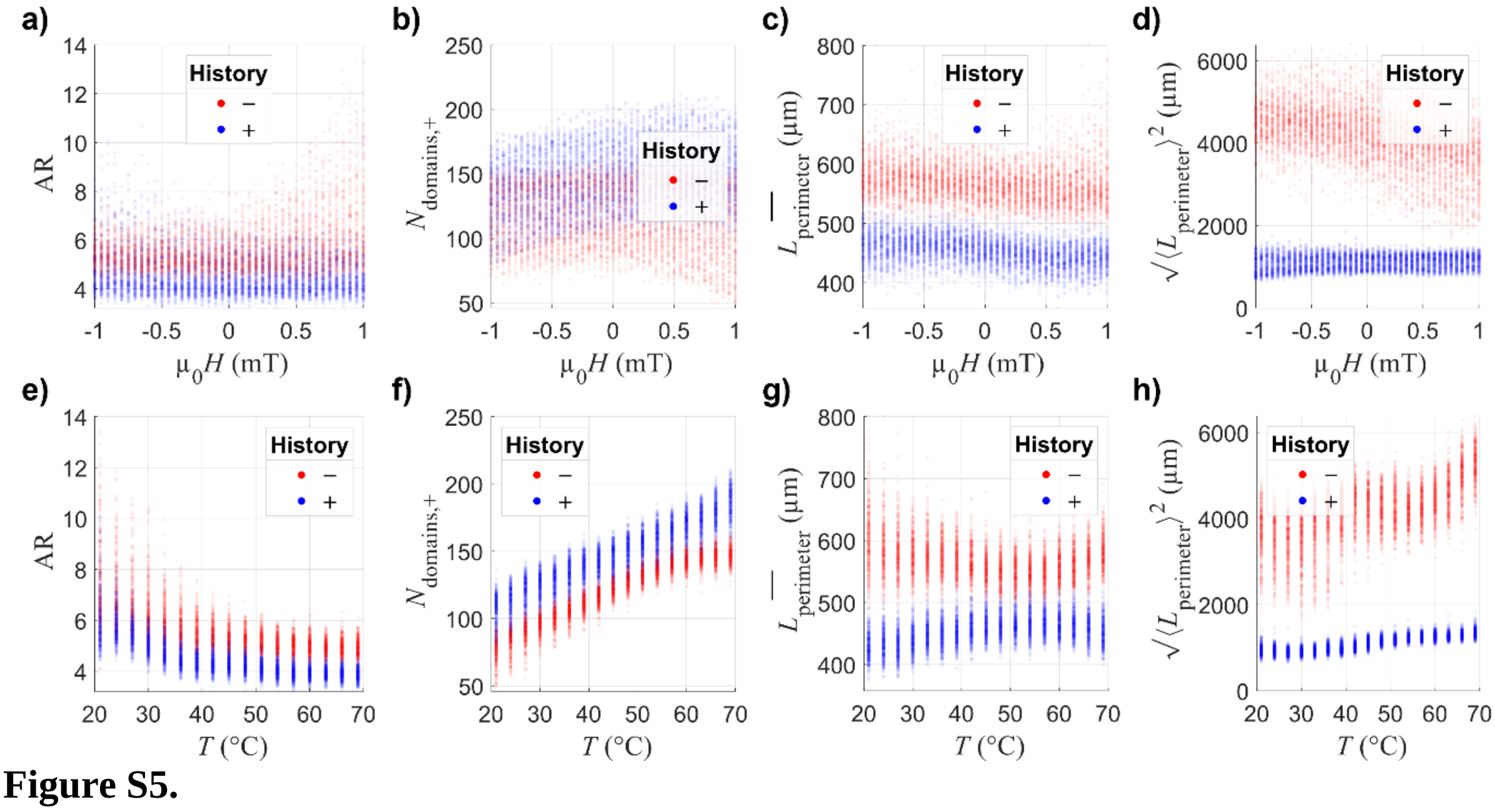

**Figure S5.**

| | |
|---|---|
| AR | **Fig. S4a,e:** Average aspect ratio of the domains major and minor axis, indicating the elongation of the domains, or inversely how clumped up the domains are. |
| $N_{+,domains}$ | **Fig. S3b,f:** Total number of separate positive magnetic domains. The temperature dependence is directly correlated to the domain period. |
| $\overline{L_{perimeter}}$ | **Fig. S3c,g:** Average outer perimeter of connected positive magnetic domains. Equivalent to the average total length. |
| $\sqrt{\langle L_{perimeter}\rangle^2}$ | **Fig. S4d,h:** Standard deviation of outer perimeter of connected positive magnetic domains. Equivalent to the average total length. |

**Feed Forward Neural Network**

Adam optimizer with mean square error loss function
Number of epochs = 400
Learning rate = $4\cdot10^{-5}$
Weight decay = $1\cdot10^{-4}$

```
class MLPRegressor(nn.Module):
    def __init__(self):
        super().__init__()
        self.net = nn.Sequential(
        nn.Linear(20, 512),
        nn.ReLU(),
        nn.Linear(512, 1024),
        nn.ReLU(),
        nn.Dropout(0.03),
        nn.Linear(1024, 512),
        nn.ReLU(),
        nn.Linear(512, 256),
        nn.ReLU(),
        nn.Linear(256, 128),
```

```
        nn.ReLU(),
        nn.Linear(128, 64),
        nn.ReLU(),
        nn.Linear(64, 32),
        nn.Linear(32, 16),
        nn.Linear(16, 3)
        )
```

**Convolutional Neural Network**

Adam optimizer with mean square error loss function

Number of epochs = 70

learning rate = $4\cdot10^{-6}$with a gamma = 0.65 and a step size of 4.

```
class CNNModel(nn.Module):
    def __init__(self):
        super(CNNModel, self).__init__()
        self.conv1 = nn.Conv2d(1, 16, kernel_size=3, padding=1)
        self.bn1 = nn.GroupNorm(8,16)
        self.conv2 = nn.Conv2d(16, 32, kernel_size=3, padding=1)
        self.bn2 = nn.GroupNorm(8,32)
        self.pool1 = nn.MaxPool2d(2, 2) # 1024x1224 -> 512x612

        self.conv3 = nn.Conv2d(32, 64, kernel_size=3, padding=1)
        self.bn3 = nn.GroupNorm(8,64)
        self.conv4 = nn.Conv2d(64, 128, kernel_size=3, padding=1)
        self.bn4 = nn.GroupNorm(8,128)
        self.pool2 = nn.MaxPool2d(2, 2) # 512x612 -> 256x306

        self.conv5 = nn.Conv2d(128, 128, kernel_size=3, padding=1, dilation=1)
        self.bn5 = nn.GroupNorm(8,128)
        self.conv6 = nn.Conv2d(128, 128, kernel_size=3, padding=1)
        self.bn6 = nn.GroupNorm(8,128)
        self.pool3 = nn.MaxPool2d(2, 2) # 256x306 -> 128x153
        self.conv7 = nn.Conv2d(128, 128, kernel_size=3, padding=1, dilation=1)
        self.bn7 = nn.GroupNorm(8,128)

        self.adapt_pool = nn.AdaptiveAvgPool2d((16, 16)) # → 128 x 16 x 16
        self.fc1 = nn.Linear(128 * 16 * 16, 256)
        self.dropout1 = nn.Dropout(0.03)
        self.fc2 = nn.Linear(256, 3)
```

**Decoding Magnetic Texture**

**Table of Content**

Magnetic textures of magnetic domain structures are the result of external and material parameters, encoding their information in a non-linear, multivariate fingerprint in equivalent, but distinct structures. Combining quantitative magneto-optical imaging with machine learning and physically interpretable features, we decode its information about temperature, as well as past and present magnetic field, from a single image.

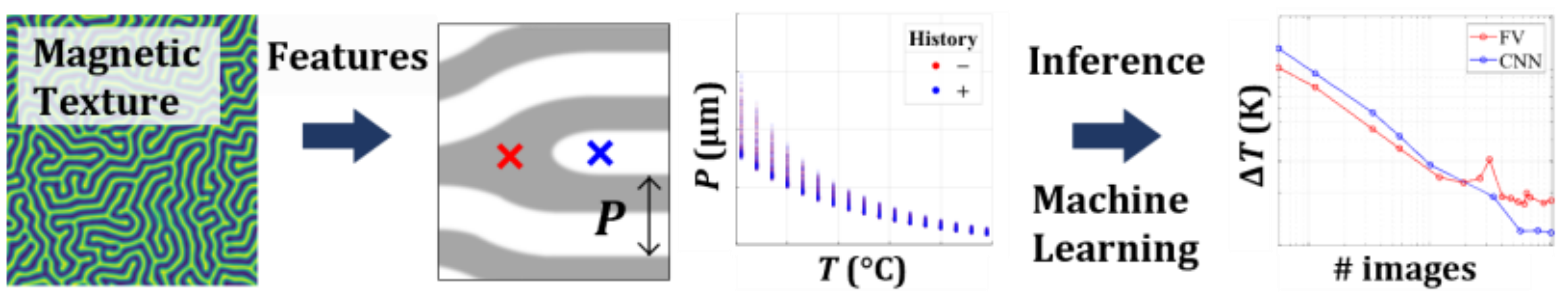